\documentclass[12pt]{article}
\usepackage{amsmath,amssymb}
\usepackage[dvips]{graphicx}
\usepackage{cite}

\def\CC{{\cal C}}

\def\CO{{\cal O}}

\newcommand{\fs}{{\bar{\cal A}}}
\newcommand{\As}{{\cal A}}

\newcommand{\bbibitem}[1]{\bibitem{#1}\marginpar{#1}}

\def\Label#1{\label{#1}%
  \smash{\hbox to0pt{\raise1ex\hbox{\tiny[#1]}\hss}}}
\def\noLabels{\let\Label=\label}
\def\nobbibitem{\let\bbibitem=\bibitem}

\newcommand{\be}{\begin{equation}}
\newcommand{\ee}{\end{equation}}
\newcommand{\bea}{\begin{eqnarray}}
\newcommand{\eea}{\end{eqnarray}}
\newcommand{\beq} {\begin{equation}}
\newcommand{\eeq} {\end{equation}}
\newcommand{\beqa} {\begin{eqnarray}}
\newcommand{\eeqa} {\end{eqnarray}}

\newcommand{\mB}{\mathcal{B}}

\newcommand{\nn}{\nonumber}
\newcommand{\eq}[1]{(\ref{#1})}
\newcommand{\inv}[1]{\frac{1}{#1}}
\newcommand{\morder}[1]{{\cal O}\left(#1\right) }
\newcommand{\pat}{\partial}

\newcommand{\zbar}{\bar{z}}

\def\eps{\epsilon}

\def\ie{\emph{i.e.}}
\def\eg{\emph{e.g.}}
\def\la{\lambda}
\def\om{\omega}
\def\appro{\mathrm{aprx}}
\def\asymp{\mathrm{asymp}}
\def\vek{\vec{k}}
\def\veX{\vec{X}}
\def\vex{\vex}

\begin{document}


\rightline{HIP-2008-18/TH}
\vskip 2cm \centerline{\Large {\bf
$N$-point functions in rolling tachyon background}}

\vskip 1cm

\renewcommand{\thefootnote}{\fnsymbol{footnote}}

\centerline{{\bf Niko Jokela,$^{1,2}$\footnote{najokela@physics.technion.ac.il} Matti
J\"arvinen,$^{3}$\footnote{mjarvine@ifk.sdu.dk}
and Esko Keski-Vakkuri$^{1,2}$\footnote{esko.keski-vakkuri@helsinki.fi} }}
\vskip .5cm \centerline{\it ${}^{1}$Helsinki Institute of Physics
and ${}^{2}$Department of Physics } \centerline{\it
P.O.Box 64, FIN-00014 University of Helsinki, Finland} \centerline{\it ${}^{3}$ Center for High Energy Physics, University of Southern Denmark} \centerline{\it Campusvej 55, DK-5230 Odense M, Denmark}

\setcounter{footnote}{0}

\renewcommand{\thefootnote}{\arabic{footnote}}

\begin{abstract}

We study $n$-point boundary correlation functions in Timelike
Boundary Liouville theory, relevant for open string multiproduction
by a decaying unstable D-brane. We give an exact result for the
one-point function of the tachyon vertex operator
and show that it is consistent with a previously
proposed relation to a conserved charge in string theory. We also discuss
when the one-point amplitude vanishes. Using a
straightforward perturbative expansion, we find an explicit
expression for a tachyon $n$-point amplitude for all $n$, however
the result is still a toy model. The calculation uses a new
asymptotic approximation for Toeplitz determinants, derived by
relating the system to a Dyson gas at finite temperature.

\end{abstract}

\newpage

\section{Introduction and Summary}

String theory contains D-branes of opposite charge, so one should be
able to understand their annihilation process. A related problem is
the decay of a single unstable brane, such as a D-brane in bosonic
string theory. A simple model for the D-brane decay describes a
process starting from the infinite past, involving a spatially
homogeneous tachyon field rolling towards the true minimum of its
effective potential \cite{SLNT,Sen:2004nf}. A basic open problem is
to calculate tree-level string scattering amplitudes in the rolling
tachyon background, corresponding to production of multiple closed
or open strings by the decaying brane.  There are both conceptual
and technical aspects to this problem. Because the background is
time dependent, there are different ways to define the notion of
vacuum and asymptotic states. A technical framework for the bosonic
homogeneous brane decay is the timelike boundary Liouville theory
(TBL) coupled to 25 free massless spacelike bosons, and the problem
of computing $n$-point correlation functions \cite{Gutperle:2003xf}.
Calculations are difficult since they involve complicated coupled
integrals and/or nonintuitive analytic continuations.


In this paper we focus on calculating boundary $n$-point functions
in TBL. The two-point function, associated to the rate of open
string pair production by a decaying brane, has been investigated
before \cite{Strominger:2002pc,Gutperle:2003xf}, and also in a
curved spacetime (AdS$_3$) in \cite{Israel:2006ip}. (Other string
production work is found in
\cite{Gaiotto:2003rm,Lambert:2003zr,Constable:2003rc,
Karczmarek:2003xm,Okuyama:2003jk,Nagami:2003yz,
Gutperle:2004be,Bergman:2004pb,Balasubramanian:2004fz,Shelton:2004ij,Jokela:2005ha}.)
A simple toy model is obtained by moving to the minisuperspace
approximation, where strings are pointlike, and the problem reduces
to a relatively simple quantum mechanical scattering problem.
Returning back to the original setup, the standard prescription is
to start from spacelike boundary Liouville theory (SBL), where the
two-point and three-point functions have known well-defined analytic
expressions \cite{Fateev:2000ik,Teschner:2000md,Ponsot:2001ng}, and
then continue to the timelike theory by $b\rightarrow i, \phi
\rightarrow iX^0$. However, the continuation must involve a
prescription to avoid the accumulation of an infinite number of
poles and zeroes which would render the answer ill defined. One way
to motivate a prescription is by aiming to make contact with the
minisuperspace analysis. This procedure gives a physically pleasant
answer, exponentially suppressed open string pair production at high
energies. However, some doubt remains, as the prescription for the
analytic continuation was not unique and some of the steps involved
are rather indirect. It is desirable to pursue alternative
approaches, they may give further support to the previous analysis
or lead to other reasonable prescriptions. Moreover, the previous
method is difficult to extend beyond the two-point function.

An alternative method to compute correlation functions in TBL was
given in \cite{Balasubramanian:2004fz}. Instead of indirect arguments, the method \cite{Balasubramanian:2004fz} is based on a straightforward perturbative expansion, and the observation that
Random Matrix Theory (RMT) \cite{mehta} techniques become applicable to the ensuing integrals. This method was successfully applied to compute the bulk-boundary function \cite{Balasubramanian:2004fz,Jokela:2005ha}. On the other hand, the same problem
was also considered by Liouville theory methods. The bulk-boundary function was
calculated in
spacelike Liouville theory in \cite{Hosomichi:2001xc}. Ref.
\cite{Fredenhagen:2004cj} then investigated the analytic continuation from spacelike
Liouville theory to timelike theory, and found a result for the bulk-boundary
function which is similar to that of \cite{Balasubramanian:2004fz}.

In the method of \cite{Balasubramanian:2004fz}, correlation functions are
related to expectation values of periodic functions (Fisher-Hartwig
symbols) in the circular ensemble of unitary matrices (CUE), also equivalent
to Toeplitz determinants of Fourier coefficients. This observation
was extended to $n$-point functions and superstrings in \cite{Jokela:2005ha}.
Alternatively, the $n$-point functions can be related to thermal
expectation values in a classical log gas of unit charges in two
dimensions, \emph{e.g.}, the Dyson gas. In \cite{Balasubramanian:2004fz}, this observation was made at
a formal level, while the problem of actually finding explicit
answers for the correlation functions still remained.

In the present paper, we use the interpretation of the correlations
functions as thermal Dyson gas expectation values, and then use
physical insight to find analytic expressions. We are able to derive
an expression for an $n$-point amplitude. The virtue of our approach
is that it is relatively straightforward, and it is powerful enough
to for a first time yield an analytic expression for an $n$-point
amplitude for all $n$. The downside is that at the moment we do not
have quantitative control of our approximations by the time we
compute the amplitude. Consequently, we do not yet know how to
compare the result with the previous one for the open string pair
creation amplitude. Nevertheless, we consider the techniques that we
have developed to be a step forward towards full control of the
scattering problem.

The paper is organized as follows. Section \ref{sec:kaksi} begins
with a review of some facts of TBL. We present some preliminary calculations
and discuss the one-point function and the vanishing one-point amplitude.
Next, we present a contour integration
trick which is powerful in summing the series expansion for the
correlation functions. We then use the one-point function formula to test a
recently proposed master formula \cite{Kraus:2002cb} for a string
theoretic definition of a conserved charge. Section
\ref{sec:Coulomb} reviews the relation of TBL to Dyson gas at finite
temperature. In Section \ref{npointsec}, we use this connection to
derive an approximation for the integrals which appear as
coefficients in the series expansion of a  $n$-point amplitude, then
use the approximation for the coefficients and the contour
integration trick of Section \ref{sec:kaksi} to derive a toy model
result for the amplitudes. Some calculational details are left in
Appendices~\ref{sec:appKRS},~\ref{exampapp}, and~\ref{sec:zeros}.


\section{Boundary amplitudes in timelike Liouville theory}\label{sec:kaksi}

Let us first review some facts to identify the problem of interest.
Full scattering amplitudes in bosonic string theory involve
contributions from the timelike $X^0$ and the 25 spacelike
directions $\veX=(X^I)$. However, as discussed in
\cite{Lambert:2003zr}, one can simplify the calculations by adopting
a gauge where the string vertex operators factorize into a form \be
V = e^{i\omega X^0}~V_{sp}(\vec{X}) \ , \ee so that all dependence
on $X^0$ is in the simple exponential factor, while $V_{sp}$
contains the more complicated polarization tensor factor and only
depends on the spacelike directions $\vec{X}$. For a homogeneous
rolling tachyon background depending only on $X^0$, all the
complications arise from contractions in the $X^0$ direction between
the background and the vertex operators, while contractions in the
spatial directions give a simple contribution. Correspondingly, the
$n$-point correlation functions in the homogeneous rolling tachyon
background factorize into a product of an $n$-point function of
$e^{i\omega_{a} X^0 (\tau_a) }$ (where the label $a=1,\ldots ,n$) in TBL, and an $n$-point
function of $V_{sp}(\vek_{a}; \veX(\tau_a))$ in the theory of free
spacelike bosons,
\begin{eqnarray}
\left\langle \prod^{n}_{a=1} e^{i\omega_{a} X^0(\tau_a)} \right\rangle_{\rm TBL}
&& \left\langle \prod^{n}_{a=1}
V_{sp}(\vek_{a};\veX(\tau_a)) \right\rangle_{\rm free} \nonumber \\
&& \mbox {} \equiv e^{-i\sum_a \vek_a \cdot \vec{x}}
\left\langle \prod^{n}_{a=1} e^{i\omega_{a} X^0(\tau_a)}
\right\rangle_{\rm TBL} F_{\rm free}[(\vek_a);(\tau_a)] \ ,
\end{eqnarray}
where we separated the spacelike zero modes. The
on-shell conditions $k_a^2 = -\omega^2_a + \vek_a^2 = -m^2_a$
can be satisfied for a range of values of $\omega_a$. The problem of
interest is to calculate $n$-point functions in TBL for generic
$\omega_a$. We will also try to compute the full scattering
amplitude for $n$ open string tachyons.

The action of TBL is
\be\label{eq:TBL}
 S_{\rm TBL} = -\frac{1}{2\pi}\int_{\rm disk} \pat X^0\bar\pat X^0 +\lambda\oint dt e^{X^0} \ .
\ee
Eventually we will be interested in the open string $n$-point tachyon amplitude
\begin{eqnarray}\label{eq:ekaampli}
\As_n(\omega_1,\vek_1;\ldots ;\omega_n,\vek_n) &&= \int d^p\vec{x} e^{-i\sum_a \vec k_a\cdot \vec x}
\int \prod^n_{a=1}\frac{d\tau_a}{2\pi} F_{\rm free}[(\vek_a);(\tau_a)] \\
&& \times \int {\cal D}X^0 e^{-S_{TBL}} \prod^n_{a=1} e^{i\omega_a X^0(\tau_a )} \ , \nonumber
\end{eqnarray}
at tree level, where the momenta $\vek_a$ are in the spacelike directions of the decaying $p$-dimensional
brane, $\tau_a$ denote points on the boundary of the disk (unit circle).\footnote{We could
use the conformal Killing group (CKG) $PSL(2,R)$ to fix three of the vertex operator coordinates $\tau_a$,
but we have chosen to leave them unfixed and average over the locations.}
For tachyons the contribution from the spacelike directions (with divergent self-contractions removed) is
\be\label{eq:F}
F_{\rm free}[(\vek_a);(\tau_a)] =
\prod_{a<b}|e^{i\tau_a}-e^{i\tau_b}|^{2\vek_a\cdot\vek_b} \ ,
\ee
with the on-shell condition $k_a^2 =
-\omega^2_a + \vek_a^2 = 1$.
The conservation of spatial momentum has been discussed,
{\em e.g.}, in \cite{Balasubramanian:2004fz}. As discussed in the Introduction,
different approaches have been used for the calculation. We will follow the approach
of \cite{SLNT,Balasubramanian:2004fz} and first expand $\As_n$ as a power series, in
powers of the boundary interaction. We also separate out the overall zero mode $x^0$ dependence, so $\As_n$ becomes
\bea
 \As_n & = & \delta_{0,\sum_a \vek_a}
 \int \prod^n_{a=1}\frac{d\tau_a}{2\pi} F[\cdots ]\int dx^0 e^{ix^0 \sum^n_{a=1}\omega_a } \sum^\infty_{N=0} \frac{(-2\pi
 \lambda e^{x^0})}{N!} \nonumber \\
  & & \times \int \prod^N_{i=1} \frac{dt_i}{2\pi} \left\langle
   e^{X'^0(t_1)} \cdots e^{X'^0(t_N)} \prod^n_{a=1} e^{i\omega_a X'^0(\tau_a)} \right\rangle
   \ .
\eea
After the Wick contractions and substituting the Green's functions, the amplitude takes the
form of a power series of coupled integrals.

The amplitude $\As_n$ then becomes
\bea\label{nptamplitude}
 && \As_n (\xi_1,\ldots ,\xi_n) =  \delta_{0,\sum_a \vek_a}
 \int dx^0 \exp\left[x^0\sum_{a=1}^n \xi_a\right]
 \fs_n(2\pi\lambda e^{x^0}) \ \ , \ \  {\rm where} \nonumber \\
&& \fs_n(z) = \sum_{N=0}^\infty(-z)^N
I_{\xi_1,\ldots,\xi_n}(N) \label{Andef} \ ,
\eea
where we have adopted the notation
\be\label{eq:zdef}
   z \equiv 2\pi \lambda e^{x^0} \ \ ; \ \  \xi_a \equiv i\omega_a \ ,
\ee
and the integrals
\beqa \label{Indef}
 I_{\xi_1,\ldots,\xi_n}(N) &=& \inv{N!} \int \prod_{i=1}^{N}\frac{dt_i}{2\pi}\prod_{a=1}^{n}\frac{d\tau_a}{2\pi}
 \left[\prod_{1\leq i<j\leq N}|e^{it_i}-e^{it_j}|^{2}\right]\nn\\
 &&\times \left[\prod_{i=1}^{N}\prod_{a=1}^{n}|e^{i\tau_a}-e^{it_i}|^{2\xi_a}\right]
 \left[\prod_{1\leq a<b\leq n}|e^{i\tau_a}-e^{i\tau_b}|^{2\xi_a\xi_b+2\vec{k}_a\cdot \vec{k}_b}\right]\ ,
\eeqa
which include the spacelike contribution $F$. In order to do the sum over $N$ we need to work 
out the $t_i$ integrals for arbitrary $N$. When calculating the integrals it is often useful 
to assume that $\xi_a$ are positive real numbers and continue to imaginary $\xi_a$, \ie, to 
real energies $\om_a$, in the end. This is not problematic since the $t_i$ integrals converge 
for $\mathrm{Re}\, \xi_a>-1/2$ and thus define an analytic function of $\xi_a$ in this region.

\subsection{Some preliminary considerations}\label{1ptsec}

The simplest case to consider is $n=1$, the one-point boundary amplitude. Invariance under translation requires the one-point function to vanish unless the operator at the boundary has zero conformal weight,
rendering the case trivial. However, it turns out that some calculations will be useful
for the nontrivial case $n>1$.
It is also known that in a noncompact conformal field theory (CFT) integrated one-point functions can be nonzero \cite{Kraus:2002cb,Seiberg:1990eb}. Ref. \cite{Kraus:2002cb} proposed a relation between a one-point
function and a spacetime boundary term. In our case, we can use a TBL one-point function as a check
of the master formula in \cite{Kraus:2002cb}, and find it to be consistent.

Let us postpone other discussions for a moment and just focus on a straightforward calculation. We consider
the series that appears in (\ref{nptamplitude}), with $n=1$,
\be \label{Adef}
 \fs_1(z) =  \sum_{N=0}^\infty (-z)^{N} \cdot I_\xi(N) \ ,
\ee
where now
\beqa \label{Idef}
 I_\xi(N) &=& \inv{N!}\int_0^{2\pi}\frac{d\tau}{2\pi}\int\left[\prod_{i=1}^{N}\frac{dt_i}{2\pi}|e^{i\tau}-e^{it_i}|^{2i\omega}\right]
 \left[\prod_{1\leq i<j\leq N}|e^{it_i}-e^{it_j}|^{2}\right] \\
 &=& \inv{N!}\int\left[\prod_{i=1}^{N}\frac{dt_i}{2\pi}|1-e^{it_i}|^{2\xi}\right]
 \left[\prod_{1\leq i<j\leq N}|e^{it_i}-e^{it_j}|^{2}\right] \ .
\eeqa
Here we denoted $\xi=i\omega$, where $\om$ is the energy of the open string.\footnote{Note that
(after removing the self-contractions in the spacelike directions) $F_{\rm free} =1$.} It is interesting
to note that the integrand is independent of $\tau$, the coordinate of the vertex operator at the boundary,
so that the $\tau$ integral is trivial.
In other words, the integrand is invariant under translations along the boundary, independently of
$\xi = i\omega$. However, the total one-point function also contains the contribution from the
spacelike directions with a $\delta_{0,k}$ factor, which along with the on-shell condition will
constrain $\omega$. But let us focus back to the properties of the series (\ref{Adef}).

The same series has been considered in the context of a general bulk-boundary
amplitude, which has been calculated in closed form in \cite{Balasubramanian:2004fz,Jokela:2005ha}.
The bulk-boundary amplitude involves
\be \label{bbamp}
 \hat\As_{1+1}(\omega_c,\omega_o) \equiv\int\! dx^0 e^{i(\om_o+\om_c) x^0} \sum_{N=0}^\infty (-z)^N I_{i\om_o}(N) \ ,
\ee
where $\omega_c$ is the energy of the bulk operator $\exp \{i\omega_cX^0 (z, \bar{z})\}$
and $I_{i\omega_o}(N)$ is the integral (\ref{Idef}) evaluated at $\xi=i\omega_o$,
where $\omega_o$ is the energy
of the boundary operator.\footnote{The one-point amplitude 
is formally the limit $\omega_c \to 0$ of the bulk-boundary amplitude (\ref{bbamp}).
We also omitted a $\delta$-function term [see Subsection \ref{sec:contour}].}
First, the integral evaluates to the relatively simple expression
\beq \label{Ires}
 I_\xi(N)=\prod_{j=1}^{N} \frac{\Gamma(j)\Gamma(j+2\xi)}{\Gamma(j+\xi)^2} = \frac{G(\xi+1)^2}{G(2\xi+1)}\frac{G(N+2\xi+1)G(N+1)}{G(N+\xi+1)^2} \ , 
\eeq
where $G$ is the Barnes $G$ function. After converting to integral representation of
the $\Gamma$ functions, the sum over $N$ in \eq{Adef} can be done \cite{Balasubramanian:2004fz,Jokela:2005ha},
leading to the result
\beqa \label{VEPAres}
 \hat \As_{1+1}(\omega_c,\omega_o)
 = -i\pi \frac{(2\pi\la)^{-i(\omega_c+\omega_o)}}{\sinh\pi(\omega_c+\omega_o)}
 \exp\left[\int_0^\infty\frac{dt(1-e^{-i\om_o t})^2}{2t(1-\cosh t)}(1-e^{i(\om_c+\om_o)t})\right] \ .
\eeqa

We would first like to point out an interesting feature, which was not
investigated in \cite{Balasubramanian:2004fz,Jokela:2005ha}. Let us write it in terms of the Barnes $G$ functions,
using the integral representation
\beq
 \log G(z+1) = \int_0^\infty \frac{dt}{t}e^{-t}
 \left[\frac{z(z-1)}{2}-\frac{z}{1-e^{-t}}+\frac{1-e^{-zt}}
 {\left(1-e^{-t}\right)^2}\right]\ \mathrm{;}\qquad \mathrm{Re}(z)>-1\ .
\eeq
We find
\be \label{bbres}
 \hat \As_{1+1}(\omega_c,\omega_o)
   =  -i\pi \frac{(2\pi\la)^{-i(\omega_c+\omega_o)}}{\sinh\pi(\omega_c+\omega_o)} J_{i\omega_o}\left( i(\omega_c+\omega_o)\right) \ ,
\ee
where
\be \label{Jdef}
 J_{\xi}\left(s\right) = \frac{G(\xi+1)^2}{G(2\xi+1)}\frac{G(2\xi-s+1)G(-s+1)}{G(\xi-s+1)^2} \ .
\ee

The asymptotic behavior \cite{Balasubramanian:2004fz,Jokela:2005ha}
follows easily from \eq{Jdef},
\beq
 J_{i\omega_o}\left( i(\omega_c+\omega_o)\right) \substack{\phantom{\om} \\ \sim \\ \om_c\to \infty}
 \om_c^{-\om_o^2} \ ,
\eeq
by using the asymptotic series of the Barnes $G$ function
\be\label{eq:Barnesasymp}
 \log G(z+1) = z^2\left(\inv{2}\log z-\frac{3}{4}\right) +\frac{z}{2} \log 2\pi -\inv{12}\log z + \zeta'(-1) + \morder{1/z^2} \ .
\ee

An interesting feature is that \eq{Jdef} is a natural continuation of \eq{Ires} to noninteger
values, replacing $N\rightarrow -s$,
but \eq{Ires} was the $N$th coefficient in the series \eq{bbamp}, while \eq{Jdef} is essentially
the sum.\footnote{A similar observation has been made in the case of bulk amplitudes in spacelike
Liouville theory in \cite{Zamolodchikov:1995aa}.} We will show 
how coefficients convert to the sum in the next Subsection \ref{sec:contour}, by a new contour integral trick which
also allows a more controlled investigation of the convergence of the series \eq{bbamp}.
The other benefit of the calculation is that it can also be applied to $n$-point amplitudes.
But let us first continue with the one-point function.

As seen by comparing \eq{Adef} and \eq{bbamp},
we can formally use the result \eq{bbres} to obtain a formula for
the Fourier transform of \eq{Adef} by setting $\om_c=0$ and $\om_o=\om$, giving
\bea \label{1ptres}
 \hat \As_1(\om)   &=& \hat \As_{1+1}(0,\om)
        \nn \\
         & = & -i\pi\frac{(2\pi\lambda)^{-i\omega}}{\sinh \pi \omega}
 \exp\left[-\int_0^\infty dt\, \frac{(1-e^{-i\omega t})(1-\cos\omega t)}{t(1- \cosh t)} \right]\nn\\
 &=& (2\pi\la)^{-i\omega} \Gamma(i\om) \frac{G(i\om+1)^3G(2-i\om)}{G(2i\om+1)} \ .
\eea
Notice that we will carefully rederive this formula in the next subsection.
The singularities and zeroes of this function are listed in Appendix~\ref{sec:zeros}. In particular,
the zeroes are located at the imaginary axis, at $\omega =in$, where $n$ is an integer, except at
$n=0,\pm 1$. Consider then the full one-point tachyon amplitude (the $n=1$ case of \eq{nptamplitude})
\be
   \As_1 (\omega ) = \delta_{0,\vek} \int dx^{0} \exp (i\omega x^0) \As_1 (2\pi \lambda e^{x^0})
           = \delta_{0,k} \hat \As_1 (\om) \ .
\ee
The momentum conservation condition $\vek=0$ along with
$\omega^2 = -1+\vec k^2$ demands $\omega=\pm i$
so that the amplitude involves the operator $\exp (\mp X^0)$. The result is
\be
 \As_1 (\omega ) =  \delta_{0,\vek} \frac{1}{2}(\pi \lambda)^{\pm 1} \ .
\ee
Note that the choice $\omega =-i$ is related to the disk partition function by
\be
  \hat\As_1 (\omega=-i) = -\frac{1}{2\pi}\int dx^{0} \frac{\partial}{\partial \lambda}Z_{\rm disk,\lambda}(x^0) = \frac{1}{2\pi \lambda} \ ,
\ee
where $Z_{\rm disk,\lambda}(x^0) = \fs_0(x^0) = 1/(1+2\pi\la  e^{x^0})$.
Conversely, for $\omega \neq \pm i$, the on-shell condition requires $\vek\neq 0$ so that the one-point amplitude vanishes.
Even though the amplitude vanishes for generic $\omega$,
the expression \eq{1ptres} will be met again in the context of higher point amplitudes.
It will be interesting
to know its  asymptotic behavior in the limit $|\om| \to \infty$.
It can be calculated to arbitrary order
by using the asymptotics of Barnes $G$ \eq{eq:Barnesasymp}.
The leading terms are
\beqa \label{1ptasympt}
 \hat\As_1(\omega) & = & -i\pi \frac{(2\pi\la)^{-i\omega}}{\sinh\pi\omega}
 \exp\left[\om^2\left(\frac{i\pi}{2}\mathrm{sgn}\left(\mathrm{Re}\,\omega\right) +2\log 2\right)
 -\frac{1}{4}\log\left(i\om\right) \right. \nn\\
& & \left.-\frac{i\pi}{12}\mathrm{sgn}\left(\mathrm{Re}\,\omega\right)+\frac{1}{12}\log 2+3\zeta'(-1) \right] \left[1+\CO\left(\om^{-2}\right)\right] \ ,
\eeqa
where $\arg(\omega) \ne \pm \pi/2$.

\subsection{A contour integral method}\label{sec:contour}

Next we calculate the integrated amplitude using a contour integration trick
which allows us to sum the series over $N$ in \eq{Andef} and analytically continue the resulting amplitude to the region where the defining sum is not convergent.
The essential required feature of the coefficients $I_\xi(N)$
is that they should not diverge too fast for large $N$. For concreteness and simplicity we will first consider the series \eq{Adef} and \eq{bbamp}. However, our method can also be applied to higher point functions
as we will discuss in Section~\ref{npointsec}. More precisely, the calculation can be generalized to the case of the $n$-point amplitude (\ref{nptamplitude}) if
we use a suitable approximate form for the integral coefficients $I_{\xi_1,\ldots ,\xi_n}(N)$.
As the contour integration method
enables us to control the convergence of the sum and the integral
it is
more rigorous than the original calculation in \cite{Balasubramanian:2004fz,Jokela:2005ha}.

We begin by studying the analytic structure of $J_\xi (s)$ of \eq{Jdef} and the asymptotics of $I_\xi(N)$ for large $N$.
We will first consider the case where $\xi$ is real and positive.
Recall the continuation of the coefficient formula \eq{Ires}
to noninteger values of $N=-s$, given by \eq{Jdef}.
From the asymptotic formula of the Barnes $G$ funtion \eq{eq:Barnesasymp}
it immediately follows that $J_\xi(s)$ has a powerlike behavior for large $s$,
\be \label{Jas}
 J_\xi(s) =\frac{G(\xi+1)^2}{G(2\xi+1)}(-s)^{\xi^2}\left[1+\morder{\inv{s}}\right] \ ; \qquad \arg s \ne 0 \ .
\ee
In addition, since $G(z+1)$ is an entire function with zeroes at $z=-1,-2,\ldots$, the poles of $J_\xi(s)$ are located at\footnote{For $\xi=1,2,\ldots$ the poles are found at $s=2\xi,2\xi+1,\ldots$.} $s=\xi+1,\xi+2,\ldots$.

\begin{figure}[ht]
\begin{center}
\noindent
\includegraphics[width=0.8\textwidth]{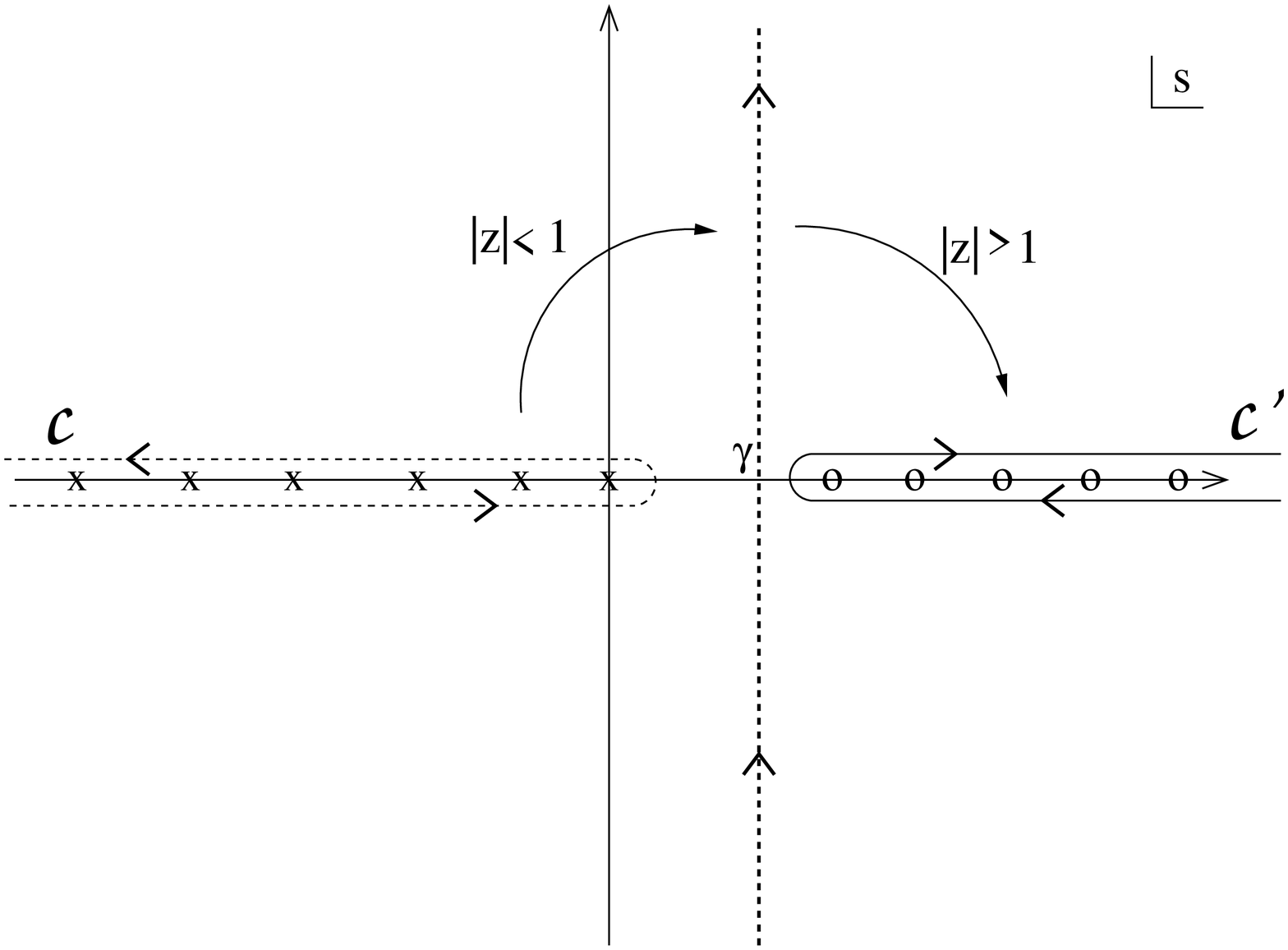}
\end{center}
\caption{The different integration contours on the $s$ plane that define the analytic continuation
of $\fs_1(z)$ for all  values of $z$.
Integration over the contours $\CC$, $\gamma+i\mathbf{R}$, and $\CC'$,
converge for $|z|<1$, $|\arg z|<\pi$, and $|z|>1$, respectively.
The x's and the o's denote the poles of $J_\xi(s)/\sin \pi s$.}
\label{epscontours}
\end{figure}

Thus in the region $|z|<1$, where the sum in
\eq{bbamp} converges, the asymptotic behavior of $J_\xi$ in \eq{Jas} enables us to write the sum as
\be \label{gasint}
 \fs_1(z)=\sum_{N=0}^\infty (-z)^{N}
 \cdot I_\xi(N)  = \inv{2\pi i} \oint_\CC \frac{\pi z^{-s}}{\sin \pi s} J_\xi(s)\,ds \ ,
\ee
where the contour $\CC$ wraps around the negative real $s$ axis as depicted
in Figure~\ref{epscontours}, picking up the residues at the poles of $1/\sin (\pi s)$
at $s=0,-1,-2,\ldots$ which produce the terms in the series. Note that
the zeroes of $G(1-s)$ in $J_\xi(s)$ cancel the poles of $1/\sin (\pi s)$ for $s=1,2,3,\ldots$.
Since $1/\sin(\pi s)$ vanishes exponentially for large imaginary $s$, we may deform
the contour (keeping $|z|<1$) in \eq{gasint} to
\be \label{invLapl}
 \fs_1(z) = \inv{2\pi i}\int_{\gamma-i\infty}^{\gamma+i\infty} \frac{\pi z^{-s}}{\sin \pi s} J_\xi(s)\,ds \ ,
\ee
where $0<\gamma<\xi+1$. This integral converges everywhere except for
negative real $z$ (if the principal branch of $z^{-s}$ with $|\arg z|<\pi$ is used)
and thus defines the analytic continuation of $\fs_1(z)$ to $|z|\ge 1$, $|\arg z|<\pi$.
Moreover, for $|z|>1$ we can continue to deform the contour to
\be \label{Cpint}
 \fs_1(z)=\inv{2\pi i} \oint_{\CC'} \frac{\pi z^{-s}}{\sin \pi s} J_\xi(s)\,ds \ ,
\ee
where $\CC'$ wraps around the positive real $s$ axis as shown in Figure~\ref{epscontours}. The integral is convergent for all $|z|>1$ so there are no singularities in this region but a logarithmic branch cut ending at $z=\infty$ which arises from the factor $z^{-s}$. The residue contributions at the poles of $J_\xi(s)$ at $s=\xi+1,\xi+2,\ldots$ give the $1/z$ expansion
\be \label{lzas}
 \fs_1(z) = (C_\xi+D_\xi\log z)z^{-\xi-1}\left[1+\morder{\inv{z}}\right] \ ,
\ee
where the constants $C_\xi,D_\xi$ can be calculated using \eq{Jdef}.

To summarize,
from the different integral representations \eq{gasint}, \eq{invLapl}, and \eq{Cpint} it follows that the only singular points of $\fs_1(z)$ are $z=-1$ and $z=\infty$.
In particular, on the integration path in the one-point amplitude 
\be \label{1ptdef}
 \hat \As_1(\xi) = \int_{-\infty}^\infty dx^0 e^{\xi x^0}\fs_1(x^0)
\ee
\ie , $z=2\pi\la e^{x^0}=0 \ldots\infty$,
$\fs_1(z)$ has no singularities. Using the series in \eq{gasint} and in \eq{lzas} we see that the integrand vanishes
exponentially
\bea
 e^{\xi x^0} \fs_1(x^0) &\substack{\phantom{so}\\ \sim \\ x^0 \to \infty}& e^{-x^0} \nn \\
 e^{\xi x^0} \fs_1(x^0) &\substack{\phantom{so}\\ \sim \\ x^0 \to -\infty}& e^{\xi x^0}
\eea
for large $\pm x^0$ so the integral over $x^0$ in \eq{1ptdef} is convergent.
Moreover, note that inserting the definition of $z$ in \eq{invLapl} we find
\be \label{invLapl2}
 \fs_1(x^0) = \int_{\gamma-i\infty}^{\gamma+i\infty} \frac{\pi(2\pi\la)^{-s}}{2\pi i}
 \frac{e^{-sx^0}}{\sin \pi s} J_\xi(s)\,ds
\ee
which defines the inverse of the (bilateral) Laplace transform. The inverse relation then gives
the master formula for the one-point amplitude 
in terms of $J_\xi (s)$,
\be \label{Lapl}
 \widetilde{\As} (s) \equiv \int_{-\infty}^\infty dx^0 e^{sx^0}\fs_1(x^0)=\frac{\pi(2\pi\la)^{-s}}{\sin \pi s}J_\xi(s) \ .
\ee
The steps from (\ref{gasint}) to (\ref{Lapl}) show how the analytic continuation of
the coefficients of the series ends up as its sum.
From the asymptotics of $\fs_1(x^0)$ we see that \eq{Lapl} converges for $0<s<\xi+1$, as expected from
the positions of poles of $J_\xi(s)/\sin \pi s$ [and the choice of $\gamma$ in \eq{invLapl}].
In particular,
\be \label{Aresapp}
 \hat \As_1 (\xi)= \widetilde{\As} (s=\xi) =  \frac{\pi(2\pi\la)^{-\xi}}{\sin \pi \xi}J_\xi(\xi)
 =(2\pi\la)^{-\xi} \Gamma(\xi) \frac{G(1+\xi)^3 G(2-\xi)}{G(2\xi+1)}
\ee
reproducing the result \eq{1ptres} above.

In the end, we want to continue the result \eq{Aresapp} for the  integrated amplitude for
imaginary $\xi=i\omega$. For imaginary $\xi$ the above analysis is not essentially changed:
the poles of $J_\xi(s)$ move to $s=i\om+1,i\om+2,\ldots$, but still lie to the right of the
imaginary axis, so that $\fs_1(x^0)$ vanishes exponentially $\fs_1(x^0) \sim e^{-x^0}$
for $x^0 \to \infty$. However, after inserting $s=i\om$ in \eq{Lapl} the convergence in
the opposite direction $x^0 \to -\infty$ is lost. We find instead
\be
 e^{i\om x^0}\fs_1(x^0) \substack{\phantom{so}\\ \sim \\ x^0 \to -\infty} e^{i\om x^0}
\ee
which signals the presence of a $\delta$ function. Indeed, the integral can be
interpreted as\footnote{The result can be checked explicitly by
writing $\fs_1(x^0)= \fs_1(x^0)\big|_{\om=0}+\left[\fs_1(x^0)-\fs_1(x^0)\big|_{\om=0}\right]$ where the first term
is simple to integrate and the latter does not contribute to the singularity.}
\bea \label{Afinal}
 \hat\As_1 (\omega ) 
  &=& \pi \delta(\om) +(2\pi\la)^{-{i \om}} \Gamma({i \om}) \frac{G(1+{i \om})^3 G(2-{i \om})}{G(2{i \om}+1)} \nn\\
 &=& (2\pi\la)^{-{i \om}} \Gamma({i (\om-i\eps)}) \frac{G(1+{i \om})^3 G(2-{i \om})}{G(2{i \om}+1)} \ ,
\eea
where the $i\eps$ changes the value of $\om$ slightly to that direction where the $x^0$ integral is convergent.

\subsection{The one-point function as a boundary term in spacetime}\label{sec:onepoint}

As discussed in \cite{Kraus:2002cb},
one difference between CFTs in compact and noncompact
target spacetimes is that in the latter case boundary terms can spoil the
holomorphicity of the stress tensor. This modifies its OPE with
other operators, and lead \cite{Kraus:2002cb} to derive a master formula relating
the one-point function (on a sphere or at the boundary of a disk)
to a boundary term in spacetime, so as to give a string theoretic
definition for a conserved charge, as an extension from field
theory. For a disk one-point function, the master formula is
\bea\label{masfor}
 \left\langle {\cal O} (z,\bar{z})\right\rangle = &&
 {\cal \widetilde N} \int d^Dx~\partial_\mu \left\{ \int_{D_2} d^2z'
  e^{2\omega (z',\bar{z}')} \right.  \nonumber \\
  && \left[\left( \frac{z'+z}{2z} \right) (z'-z) \langle \partial X^\mu (z',\bar{z}')
  {\cal O} (z,\bar{z})\rangle_{D_2}\nonumber \right. \\
  && \left.\left. + \left( \frac{\bar{z}'+\bar{z}}{2\zbar} \right) (\zbar'-\zbar)
  \langle \partial X^\mu (z',\bar{z}') {\cal O} (z,\bar{z})\rangle_{D_2}\right] \right\} \ ,
\eea
where ${\cal \widetilde N}$ is a normalization factor, the metric on the disk is
$ds^2 = e^{2\omega (z,\bar{z})}dzdz'$ and ${\cal O}(z,\zbar)$ is a local boundary operator
in the CFT with $D$-dimensional target space.

Ref. \cite{Kraus:2002cb} considered various applications where open or
closed string background gauge fields or gravitational field were
turned on. The open string rolling tachyon background gives a nice
new nontrivial
example to test the master formula (\ref{masfor}). The
worldsheet action is nonpolynomial, and the master formula
involves two-point functions in the interacting theory. We choose the
local boundary operator to be the exponential, ${\cal O}=\exp \{i\omega X^0\}$,
inserted at\footnote{The one-point function is eventually
independent of the location.} $z=e^{i\tau}$.
Its one-point amplitude is (\ref{1ptres}), which already is a (space)time
integral. So we need to show that the integrand $\fs_1(x^0)$ can be rewritten as a total
derivative as in (\ref{masfor}). Our convention for the metric of the disk is $ds^2 = dwd\bar{w}$,
so the relation to check is
\be\label{roltachfor}
 \fs_1(x^0) = \frac{\partial \mB (x^0)}{\partial x^0} \ ,
\ee
where
\beqa
 \mathcal{B}(x^0) & = & {\cal \widetilde N} \int_\mathrm{disk} d^2w \left[\frac{w^2-e^{2i\tau}}
 {2e^{i\tau}}\langle \partial X^0(w,\bar w)
 e^{i\omega X^0(\tau)}\rangle'_{TBL}\right.\nn\\
            &   & +\left.\frac{\bar w^2-e^{-2i\tau}}{2e^{-i\tau}}
            \langle \bar \partial X^0(w,\bar w)
            e^{i\omega X^0(\tau)}\rangle'_{TBL} \right] \ ,
\eeqa
where the primes indicate that we have separated the zero mode $x^0$. To show that this relation
holds, we evaluate the right-hand side. The details of this calculation are relegated to
Appendix \ref{sec:appKRS}, in part because they involve a step that is discussed in the next Section \ref{sec:Coulomb}. The end result is that \eq{roltachfor} holds, so that the one-point function is consistent with the
general expectation from \eq{masfor}.


\section{On Coulomb gas relation}\label{sec:Coulomb}

The TBL is related to
a statistical mechanical system, the Dyson gas
of particles on a unit
circle \cite{Dyson:1962es,Balasubramanian:2006sg}.\footnote{The analogy has recently
been extended to full S-brane
(or timelike boundary sine-Gordon theory) \cite{Jokela:2007wi}
and to non-BPS half S-brane \cite{Hutasoit:2007wj}.}
The key property is that the two-dimensional Green's functions can be
interpreted as coming from two interacting Coulomb gas particles confined on a circle,
\be
 V(e^{i t_i},e^{i t_j}) = -\log|e^{it_i}-e^{it_j}| \ ,
\ee
where $t_i, t_j$ are the respective angles. The perturbation expansion in $\lambda$ of (\ref{eq:TBL})
becomes related to the grand canonical ensemble of
unit charges on the circle,
\be
 Z_{\rm G} = \sum_{N=0}^\infty
\frac{z^N}{N!} \int \left[\prod_{i=1}^N\frac{dt_i}{2\pi}\right] e^{-\beta H} \ ,
\ee
where the inverse temperature is fixed to $\beta=2$, $z$ is the fugacity,
and $N!$ accounts for identical particles. The Hamiltonian contains only a potential energy
term\footnote{See the discussion on the physical interpretation in the original
paper by F. Dyson \cite{Dyson:1962es}.}
\be
 H = \sum_{\rm pairs}V(t_i,t_j) = -\sum_{1\leq i<j\leq N}\log|e^{it_i}-e^{it_j}| \ .
\ee

In this paper we focus only on the canonical ensemble.
As discussed in \cite{Balasubramanian:2006sg}, correlators in TBL are related to adding additional particles into the ensemble. The one-point function (\ref{Adef}) requires one additional particle
with an arbitrary charge $\xi$ at an angle $\tau$. The Hamiltonian becomes
\be
 H_\xi = -\sum_{1\leq i<j\leq N}\log|e^{it_i}-e^{it_j}| -\xi\sum_{i\leq i<j<N}\log|e^{i\tau}-e^{it_i}| \ ,
\ee
and the canonical partition function is
\bea
 Z_\xi & = & \frac{1}{N!}\int\frac{d\tau}{2\pi}\int\left[\prod_i\frac{dt_i}{2\pi}\right]e^{-\beta H_\xi} \\
 & = & \frac{1}{N!}\int\frac{d\tau}{2\pi}\left[\prod_i\frac{dt_i}{2\pi}\right]
 \prod_{i<j}\left|e^{it_i}-e^{it_j}\right|^2 \prod_i\left|e^{i\tau}-e^{it_i}\right|^{2\xi} \ .
\eea
The integrand does not depend on the angle $\tau$, hence it can be consistently set to zero.
We recognize $Z_\xi = I_\xi(N)$ of (\ref{Idef}).

\begin{figure}[ht]
\begin{center}
\noindent
\includegraphics[width=0.9\textwidth]{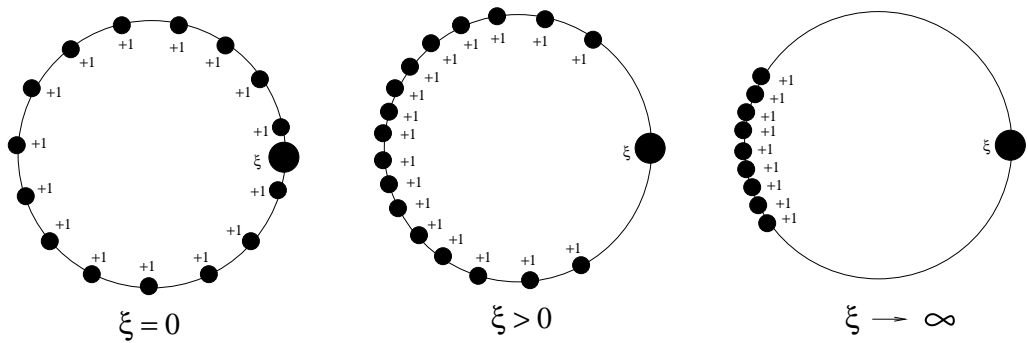}
\end{center}
\caption{Depicted is the interpretation of (\ref{gasintegral}). On the unit circle, embedded in a heat bath,
there are $N$ positive unit charges and an additional positive charge $\xi$.
As the charge strength $\xi$ increases, the repulsive force acting on the unit charges wins
over their mutual repulsion, forcing
the unit charges closer to each other on the other side of the circle.}
\label{epsfig}
\end{figure}

We can now draw insight from
the physical interpretation to better understand
the integrals and their various extensions.
As an example, consider the integral corresponding to the
canonical ensemble expectation value
\beq\label{com}
\left\langle \sum^N_{i=1} \cos(\tau-t_i) \right\rangle_{\rm can.}\!\! \equiv
 \inv{Z_\xi}\cdot\inv{N!}\!\int\! \prod_i\frac{dt_i}{2\pi} \prod_{i<j}\left|e^{it_i}-e^{it_j}\right|^2
 \prod_i\left|e^{i\tau}-e^{it_i}\right|^{2\xi}\!
  \sum_i\cos(\tau-t_i) \ ,
\eeq
which corresponds to the sum of the projected relative distances of the original charges
to the additional charge. In part by inspired guesswork we have found a result
\beq\label{gasintegral}
\left\langle \sum^N_{i=1} \cos(\tau-t_i) \right\rangle_{\rm can.}
  = -\frac{N\xi}{N+\xi}
\eeq for the integral. We have not constructed a proof for this
formula, but have checked it for $\xi=0,1,2,3,4$ and for any $N$,
and a consistency check will be given in
Appendix~\ref{exampapp}.\footnote{After the first version of this
work was finished, we were informed by
H.~Schomerus that he has constructed a proof
\cite{Schomerus:2008je}
of this 
formula. We thank him for bringing this to our attention.} We can
visualize the $\xi \rightarrow \infty$ limit (at finite $N$)  of the
result (\ref{gasintegral}) easily in Figure~\ref{epsfig}: as the
additional charge becomes stronger, it forces the unit charges
further towards the antipodal point of the circle.

\section{The $n$-point boundary amplitude}\label{npointsec}

The full $n$-point amplitude \eq{nptamplitude} is very complicated, and so are the integral coefficients \eq{Indef}
even at small $N,n \ (>1)$.
In this Section we will consider an approximation or a toy model version of a full
calculation. We begin by studying the integrals \eq{Indef}. We
interpret them as Toeplitz
determinants. One can then consider a known approximation in the large $N$ limit, and
try to improve it to be good enough to be used in the series expansion \eq{nptamplitude} at
every $N$, while hoping for it to be simple enough so that the series can be summed.
We use the Coulomb gas analogue
and find a physically motivated improved asymptotic approximation of \eq{Indef}.
This approximation agrees with the previously known asymptotics at leading order in $1/N$, but reproduces the next-to-leading $1/N$ corrections to the asymptotics of the integrals better than the old result (but not exactly). Even more importantly, it is found to work well for small values of $N$ even up to $N=0$, which contribute significantly in the final amplitude in the end.
However, the approximation is still simple enough to sum the series in $N$ to calculate the integrated $n$-point amplitude. In the approximation, essentially the ``interactions''
between the $\xi_a$ insertions can be neglected. In our end result, the $n$-point amplitude factorizes
to a product of $n$ independent one-point amplitudes.
We also present a simple example that helps to understand and motivate
the derivation in Appendix~\ref{exampapp}.

\subsection{Large $N$ asymptotics} \label{sec:asympt}

To start with, the $t_i$ integrals in \eq{Indef} can be done \cite{Balasubramanian:2004fz,Jokela:2005ha} giving
\beq \label{Inasdet}
 I_{\xi_1,\ldots,\xi_n}(N) = \int\prod_{a=1}^{n}\frac{d\tau_a}{2\pi}
 \left[\prod_{1\leq a<b\leq n}|e^{i\tau_a}-e^{i\tau_b}|^{2\xi_a\xi_b+2\vec{k}_a\cdot
 \vec{k}_b}\right] \det T_N[f] \ ,
\eeq
where $\det T_N[f]$ is the $N\times N$ Toeplitz determinant of Fourier coefficients of the function
\be\label{FH}
 f_{\tau_1,\ldots,\tau_n}(t)=\prod_{a=1}^{n}|e^{i\tau_a}-e^{it}|^{2\xi_a}\ ,
\ee
see \cite{Balasubramanian:2004fz,Jokela:2005ha} for more details.

The determinant is too complicated to allow us to sum the series (\ref{Andef}).
However, Toeplitz determinants are known to simplify at large $N$.
In particular, the large $N$ asymptotics of the determinant $\det T_N[f]$ is known for (\ref{FH}).
It reads \cite{Fisher,Widom} (see also \cite{Basor})
\be
 \det T_N[f] = N^{\sum_{a=1}^n\xi_a^2} \prod_{1\le a<b \le n}|e^{i\tau_a}-e^{i\tau_b}|^{-2\xi_a\xi_b}\prod_{a=1}^n
  \frac{G(\xi_a+1)^2}{G(2\xi_a+1)}\left[1+\morder{\inv{N}}\right] \ .
\ee
Moreover, the asymptotic behavior of \eq{Inasdet} factorizes,
\bea \label{asres}
 T_{\xi_1,\ldots,\xi_n}(N) &\equiv&  \det T_N[f] \prod_{1\le a<b \le n}|e^{i\tau_a}-e^{i\tau_b}|^{2\xi_a\xi_b} \nn\\
&=&  \prod_{a=1}^n N^{\xi_a^2}\frac{G(\xi_a+1)^2}{G(2\xi_a+1)}\left[1+\morder{\inv{N}}\right] \nn\\
  &=&  \prod_{a=1}^n\frac{G(\xi_a+1)^2}{G(2\xi_a+1)} \frac{G(N+2\xi_a+1)G(N+1)}{G(N+\xi_a+1)^2} \left[1+\morder{\inv{N}}\right] \nn\\
  &=&  \prod_{a=1}^n T_{\xi_a}(N) \left[1+\morder{\inv{N}}\right] \ ,
\eea
where $T_{\xi_1,\ldots ,\xi_n}(N)$ is the asymptotically $\tau_a$ independent factor of $\det T_N[f]$ and we used \eq{eq:Barnesasymp} to write the asymptotics in terms of Barnes $G$ functions (see also \eq{Jas}). Here $T_{\xi}(N)=I_\xi(N)$ is the one-point function discussed above in Section~\ref{sec:kaksi}.

The asymptotic formula \eq{asres} has a nice physical interpretation in terms of the
classical Coulomb gas on a circle, where $T_{\xi_1,\ldots,\xi_n}(N)$ is 
the partition function for $N$ identical unit charges at the inverse temperature $\beta=2$,
with $n$ additional particles having charges $\xi_1,\ldots,\xi_n$ at fixed angles $\tau_1,\ldots, \tau_n$.
Let us assume for a moment that all $\xi_a$ are positive integers.\footnote{This is potentially a
dangerous assumption, since eventually we want to set $\xi_a = i\omega_a$ where
typically $\omega_a$ are real, and naive continuation from integers to complex plane is
known to be problematic -- see, {\em e.g.}, the discussion in \cite{Balasubramanian:2004fz}.
We will return to this issue in the end of the Section.} Then
each particle with charge $\xi_a$ can be thought to be a cluster of $\xi_a$ unit charges.
We can then imagine constructing a typical configuration of the gas with the $n$ test charges,
from a gas of $N+\sum_a\xi_a$ unit charges, by clustering unit charges at distinct locations to form the test charges $\xi_a$. For $n< \sum^n_{a=1} \xi_a \ll N$, the typical separation of
unit charges is $\sim 1/N$, much less than the typical separation between the test charges/charge clusters. Now, we can first interpret
the $N^{\xi_a^2}$ factors in \eq{asres} arising from the self-energies of the charge clusters.
For a cluster with charge $\xi_a$, the self-energy is given by\footnote{In fact, one obtains
$E_\mathrm{self}\simeq \xi_a(\xi_a-1) \log N /2$, but the term $-\xi_a \log N/2$ cancels
against a change in the $1/N!$ factors which are discussed below.}
\be
 E_\mathrm{self}=-\sum_{1\le i<j \le \xi_a} \log |x_i-x_j|~\simeq  \sum_{1\le i<j \le \xi_a}
 \log N~\simeq \frac{\xi_a^2}{2} \log N
\ee
giving the contribution
\be
 e^{2E_\mathrm{self}} \sim N^{\xi_a^2}
\ee
to the partition function.
Second, the factorization of \eq{asres} can be understood as the absence of intercluster interactions at this
level of approximation. A heuristic argument
could be the following.
Consider a large number of unit charges on the real axis (a piece of the unit circle after magnification)
with a typical separation $d \simeq 2\pi/N$. Choose $\xi \ll N$ charges at $x_1,\ldots, x_\xi$
near the origin (so that $x_\xi \sim d$)
and perturb their locations, $x_k\rightarrow x_k-\delta_k$, by $\delta_k \sim d$ symmetrically
such that $\sum_{k=1}^\xi\delta_k=0$,
to create a cluster of charge $\xi$.
The change in the electrostatic potential after creating the cluster is then
\be
 \Delta V(x) = -\sum_{k=1}^\xi \log\left[1+\frac{\delta_k}{x-x_k}\right]
\ee
as felt at point $x$  outside the cluster, $x>x_k$. For $x\gg d$ we find that the deformation of the potential
vanishes rapidly,
\be
   \Delta V \sim \xi \left( \frac{d}{x} \right)^2 \ ,
\ee
and the contribution to the change in the total energy
in the leading order must thus come from the interaction of the cluster between the unit charges within the
region $x \sim d \sim 1/N$. However, at the distance to the neighboring clusters, the change is negligible,
so the intercluster interactions are suppressed.

\bigskip

The analysis suggests a natural way to try to improve the asymptotic formula \eq{asres}.
As explained above, in \eq{asres} intercluster interactions are absent.
However, the $1/N$ corrections due to these interactions can be added almost completely in a very simple manner.
Naturally,
each of the $\xi_a$ charges must feel the Coulomb force of the $N$ unit charges {\em and}
the other clusters with $\xi_b$ charges, $a \ne b$.
An easy modification of \eq{asres} to accommodate these is the following.  Increase the number of background unit charges acting on the $\xi_a$ cluster
from $N$ to $\tilde{N}_a = N+\sum_{b\neq a} \xi_b$
in the asymptotic formula \eq{asres}.
This indeed replaces the Coulomb force of each $\xi_b$ charge as a cluster of $\xi_b$ separate unit charges. The total effective number of unit
charges in the gas then becomes $\tilde N = \tilde{N}_a + \xi_a = N + \sum_{a=1}^n\xi_a $. We
write the improved asymptotic formula for
a renormalized Toeplitz determinant $\hat T_{\xi_1,\ldots,\xi_n}(N)$, which is simply related to $T_{\xi_1,\ldots,\xi_n}(N)$ of \eq{asres}. The modification is needed
since  $T_{\xi_1,\ldots,\xi_n}(N)$ contains the normalization
factor $1/N!$ which we want to
be replaced by $1/\tilde{N}!$:
\bea \label{normIdef}
 \hat T_{\xi_1,\ldots,\xi_n}(N) &=& \frac{N!}{\Gamma(\tilde N+1)} T_{\xi_1,\ldots,\xi_n}(N)\\
 &=& \inv{\Gamma(\tilde N+1)} \left[\prod_{1\leq a<b\leq n}|e^{i\tau_a}-e^{i\tau_b}|^{2\xi_a\xi_b}\right] \int \prod_{i=1}^{N}\frac{dt_i}{2\pi}
 \left[\prod_{1\leq i<j\leq N}|e^{it_i}-e^{it_j}|^{2}\right]\nn \\
 &&\times \left[\prod_{i=1}^{N}\prod_{a=1}^{n}|e^{i\tau_a}-e^{it_i}|^{2\xi_a}\right]
  \nn \ .
\eea

Following the discussion above, we replace the asymptotic formula \eq{asres} by an improved
formula for 
\eq{normIdef},
\be\label{newasy}
  \hat T_{\xi_1,\ldots,\xi_n}(N) =  \prod_{a=1}^n  \hat T_{\xi_a}(\tilde{N}_a)
  \left[1+\morder{\inv{N}}\right] \ ,
\ee
where
\bea
\hat T_{\xi_a}(\tilde{N}_a ) &=&
 \frac{\Gamma(\tilde{N}_a+1)}{\Gamma(\tilde{N}+1)}
 T_{\xi_a}(\tilde{N}_a) \nn \\
&=& \inv{\Gamma(\tilde{N}+1)}
\int \prod_{i=1}^{\tilde{N}_a}\frac{dt_i}{2\pi}
\left[\prod_{1\leq i<j\leq \tilde{N}_a}
|e^{it_i}-e^{it_j}|^{2}\right]
\left[\prod_{i=1}^{\tilde{N}_a}|1-e^{it_i}|^{2\xi_a}\right]
\eea
is the properly normalized partition function for a $\xi_a$ charge in the background of $\tilde N_a$ unit charges.
Inverting the relation \eq{normIdef}, we can rewrite (\ref{newasy}) as an
improved approximation for $T_{\xi_1,\ldots,\xi_n}$, 
\bea \label{gasguess2}
  T_{\xi_1,\ldots,\xi_n}(N) &\approx& T_\mathrm{norm}\prod_{a=1}^n T_{\xi_a}(\tilde{N}_a)
    \\\nn
 &=& T_\mathrm{norm}\prod_{a=1}^n \frac{G(\tilde N+\xi_a+1)G(\tilde N-\xi_a+1)}{G(\tilde N+1)^2}\frac{G(\xi_a+1)^2}{G(2\xi_a+1)} \equiv T^\appro_{\xi_1,\ldots,\xi_n}(N) \ ,
\eea
where we introduced the notation $T^\appro_{\xi_1,\ldots,\xi_n}(N)$ for
the improved asymptotics and
the normalization
factor reads
\be \label{eq:Inormdef}
 T_\mathrm{norm} =  \frac{\prod_{a=1}^n\Gamma(\tilde{N}_a +1)}{\Gamma(\tilde{N}+1)^{n-1}N!} \ .
\ee
Note that $T^\appro_{\xi_1,\ldots,\xi_{n}}$ reduces to \eq{asres} for $N \to \infty$ and still has $1/N$ corrections,
but they are expected to be essentially smaller than for \eq{asres}. In Appendix~\ref{exampapp} we discuss the simplest nontrivial example  $(\xi_1,\xi_2)=(2,2)$, where the exact results are known \cite{Jokela:2007dq,Jokela:2007yc} and find that the improved asymptotics \eq{gasguess} reduces the deviation from the exact result by more than an order of magnitude at large $N$. The new asymptotics continues to be a very good approximation to the exact result even for small values of $N$.
Moreover, note that setting, \eg, $\xi_n=1$ in \eq{gasguess} correctly reproduces $T^\appro_{\xi_1,\ldots,\xi_{n-1}}(N+1)$.

Finally, we 
collect our results in a new asymptotic approximation for the
Toeplitz determinant:
\bea
 \det T_N[f] & \approx & \prod_{1\le a<b \le n}|e^{i\tau_a}-e^{i\tau_b}|^{-2\xi_a\xi_b}
 \frac{\Gamma(N+\sum_a \xi_a +1)}{\Gamma (N+1)} \nn \\
  & & \times\prod_{a=1}^n
 \frac{\Gamma (N-\xi_a+\sum_b \xi_b +1)}{\Gamma (N+\sum_b \xi_b +1 )}
  \frac{G(\xi_a+1)^2}{G(2\xi_a+1)} \nn\\
  & & \times \frac{G(N+\xi_a+\sum_b \xi_b+1)G(N-\xi_a+\sum_b \xi_b+1)}{G(N+\sum_b \xi_b+1)^2} \ .
\eea

If we substitute this to \eq{Inasdet}, we note that
the integrals over $\tau_a$ 
give
\bea\label{eq:Inorm}
 I_{\xi_1=0,\ldots,\xi_n=0}(N) &=&
\int \prod_{a=1}^{n}\frac{d\tau_a}{2\pi}
\left[\prod_{1\leq a<b\leq n}|e^{i\tau_a}-e^{i\tau_b}|^{2\vec{k}_a\cdot \vec{k}_b}\right]
\equiv {\cal N}\left[(\vek_a )\right]  \ ,
\eea
and the result for the integral becomes
\beqa \label{gasguess}
 I_{\xi_1,\ldots ,\xi_n}(N) &\approx&
 I^\appro_{\xi_1,\ldots ,\xi_n}(N) \nn\\
&\equiv&  {\cal N}\left[(\vek_a )\right] \frac{\Gamma(N+\sum_a \xi_a +1)}{\Gamma (N+1)} \nn \\
  & & \times\prod_{a=1}^n
 \frac{\Gamma (N-\xi_a+\sum_b \xi_b +1)}{\Gamma (N+\sum_b \xi_b +1 )}
  \frac{G(\xi_a+1)^2}{G(2\xi_a+1)} \nn\\
  & & \times \frac{G(N+\xi_a+\sum_b \xi_b+1)G(N-\xi_a+\sum_b \xi_b+1)}{G(N+\sum_b \xi_b+1)^2} \ .
\eeqa
The $k_a^I$ dependence thus completely factorizes into the normalization factor $\cal N$.

There is, however, a caveat in the above derivation: the result \eq{gasguess} only makes sense when the normalization integral ${\cal N}$ is convergent. From the definition (\ref{eq:Inorm}) we see that the integral is singular whenever any of the products $\vec{k}_a\cdot \vec{k}_b \to -1/2$, which can easily occur for physical momentum values. These singularities are unphysical and they are absent in the original integral of \eq{Indef}. What happens is that for $\vec{k}_a\cdot \vec{k}_b \to -1/2$ the $\tau$ integrals become heavily peaked at $\tau_a \simeq \tau_b$. More precisely, the dominant contribution to the integral comes from the region where $\tau_b-\tau_a \sim 1/N$. In this region the large $N$ limit does not reproduce the $\tau$ dependence correctly: for positive $\xi_a\xi_b$ the integrand vanishes more rapidly than suggested by the large $N$ limit as $\tau_a-\tau_b\ \to \ 0$, which creates a cutoff for the normalization integral ${\cal N}$.

To avoid this caveat we shall assume 
that $\vec{k}_a\cdot \vec{k}_b > -1/2$ which can be satisfied together with momentum conservation only if all spatial momenta are small. Even when this condition is not met, the result \eq{gasguess} may work as a reasonable model for the $\xi$ and $N$ dependencies of \eq{Inasdet}.
We are planning to study the $\tau$ dependence of the Toeplitz determinant more closely in a forthcoming publication.

\subsection{A model amplitude}

Let us now study what can be said about the integrated amplitude
\begin{eqnarray} \label{Anidef}
  \As_n &=& \delta_{0,\sum_a \vek_a}
  \int dx^0 \exp \left[ x^0\sum_{a=1}^n \xi_a\right]\fs_n(2\pi\lambda e^{x^0}) \\
  \mbox{} &=& \delta_{0,\sum_a \vek_a}\int dx^0\exp\left[x^0\sum_{a=1}^n \xi_a\right] \sum_{N=0}^\infty(-z)^N I_{\xi_1,\ldots,\xi_n}(N) \nonumber
\end{eqnarray}
based on the asymptotic formula
\eq{gasguess}. Notice that
since the coefficients $I_{\xi_1,\ldots,\xi_n}(N)$ are asymptotically equal to a product of one-point functions they also exhibit a powerlike behavior for large $N$ [see \eq{Jas}].
This fact strongly suggests that the analysis of Subsection~\ref{sec:contour} can be extended to higher point functions, which requires that there is an analytic continuation $J_{\xi_1,\ldots,\xi_n}(s)$ of $I_{\xi_1,\ldots,\xi_n}(N)$ to complex values of $N=-s$ that has a powerlike behavior for $s\to \infty$ in all sectors of the complex $s$ plane. At least, as we shall see below, the continuation exists for the asymptotic formula \eq{gasguess} (and also \eq{asres}).
Also, we calculated $I_{\xi_1,\ldots,\xi_n}(N)$ for sets of small positive integers $\xi_a$
and for ${\vec k}_a\cdot {\vec k}_b=0$
in \cite{Jokela:2007dq}, and found that they are, in fact, polynomials of $N$. See also
Appendix~\ref{exampapp} where we treat a simple case, $(\xi_1,\xi_2)=(2,2)$, as an example.

This motivates us to check what is the result if one simply inserts the improved asymptotic formula \eq{gasguess} to \eq{Andef} and to repeat the analysis of Subsection \ref{sec:contour}.
For the sake of concreteness, we discuss the two-point function\footnote{For the two-point function spatial momentum conservation and on-shell conditions actually fix $\om_1=\om_2$.}.
As explained above, it is required that there is such an analytic continuation of $I^\appro_{\xi_1,\xi_2}(N)$ of \eq{gasguess} to complex $s=-N$ that does not blow up exponentially for $|s| \to \infty$. Remarkably,
the simplest continuation of \eq{gasguess} works:
\bea \label{Jappdef}
 J^\appro_{\xi_1,\xi_2}(s)&= &\frac{{\cal N}\left[{\vec k}_1\right]}{\Gamma(1-s)\Gamma(\xi_1+\xi_2-s+1)}\nn\\
 &&\times\prod_{a=1}^2\frac{G(\xi_a+1)^2}{G(2\xi_a+1)}\frac{G(-s+\xi_1+\xi_2+\xi_a+1) G(2-s+\xi_a)}{G(-s+\xi_1+\xi_2+1)^2}
\eea
indeed has a powerlike asymptotic behavior for large $|s|$ as can be verified using the formulae \eq{gasguess} and \eq{asres} above. Further, we need to check that the singularities of $J^\appro_{\xi_1,\xi_2}(s)$ do not
conflict with the contour deformations of Subsection~\ref{sec:contour}.
If $\xi_{1,2}>0$ the poles of $J^\appro_{\xi_1,\xi_2}(s)$ are located at $s=\xi_1+\xi_2+1,\xi_1+\xi_2+2,\ldots$. As for the one-point amplitude in Subsection \ref{sec:contour}, they are to the right of $s=\xi_1+\xi_2$, where we will evaluate $J^\appro_{\xi_1,\xi_2}(s)$ in the end [see \eq{Aguess} below]. As discussed in Subsection~\ref{sec:contour}, this means that the model two-point function
\be \label{tildeg}
 \fs^\appro_2
(z)=\sum_{N=0}^\infty (-z)^{N}I^\appro_{\xi_1,\xi_2}(N)
\ee
has no singularities for $z>0$ and vanishes sufficiently fast for $x^0 \to \infty$ to make the integral over $x^0$ in \eq{Anidef} convergent. Notice that this is not the case if the ``naive'' asymptotic formula \eq{asres} is used instead of \eq{gasguess}.\footnote{However, also the naive formula leads to a well-defined integral for imaginary $\xi_a=i\om_a$ which we will need in the end.} In particular, as discussed in Appendix~\ref{exampapp}, \eq{tildeg} has the correct asymptotic behavior for $z \to +\infty$ for integer $(\xi_1,\xi_2)$ only if one uses \eq{gasguess}.

The above checks ensure that following the analysis in  Subsection~\ref{sec:contour}
(see equations~\eq{Lapl}, \eq{Aresapp}, and \eq{Afinal})
we can sum and integrate the approximated integrals $I_{\xi_1,\xi_2}^\appro$ of \eq{gasguess}. The result is a model two-point amplitude,
\bea \label{Aguess}
 \As_2  &\approx& \delta_{0,\vek_1+\vek_2} 
\frac{-i\pi(2\pi\la)^{-i(\om_1+\om_2)}}{\sinh\pi(\om_1+\om_2)}   J^\appro_{i\om_1,i\om_2}(i\om_1+i\om_2)\nn\\
  &=& \delta_{0,\vek_1+\vek_2} {\cal N}\left[\vek_1\right](2\pi\la)^{-i(\om_1+\om_2)} \Gamma(i\om_1+i\om_2)\prod_{a=1}^2\frac{G(i\om_a+1)^3G(2-i\om_a)}{G(2i\om_a+1)} \ ,
\eea
where we already rotated to imaginary  $\xi_a=i\om_a$
and omitted a $\delta$-term. Notice the similarity to \eq{1ptres} which stems from the factorized form of the asymptotics \eq{gasguess}.

Similarly as the one-point amplitude in
\eq{Afinal}, the final result \eq{Aguess} is expected to include a term $\propto \delta(\om)$. The delta term arises in the $x^0$ integration of $\fs_2(x^0)$ from the oscillations in the region $x^0 \to -\infty$: indeed, for imaginary $\xi = \xi_1+\xi_2 = i\om_1+i\om_2$ the integrand $e^{i(\om_1+\om_2)x^0}\fs_2(x^0)$ continues to vanish exponentially for $x^0 \to +\infty$, but for $x^0 \to -\infty$ the function $\fs_2$
approaches a constant, 
which leads to oscillating behavior.
The resulting $\delta$ contribution can be isolated as follows. Write
\bea
  \As_2 &=&  \delta_{0,\vek_1+\vek_2}\int dx^0 e^{i(\om_1+\om_2)x^0} \fs_2^\appro(x^0) \nn\\
  &=& \delta_{0,\vek_1+\vek_2}\Bigg\{ I^\appro_{i\om_1,i\om_2}(N=0) \int dx^0 e^{i(\om_1+\om_2)x^0}  \fs_1(x^0)\big|_{\om=0} \nn\\
  &&+  \int dx^0 e^{i(\om_1+\om_2)x^0}  \left[\fs_2^\appro(x^0)- I^\appro_{i\om_1,i\om_2}(N=0) \fs_1(x^0)\big|_{\om=0}\right]\Bigg\}
\eea
where $\fs_1(x^0)\big|_{\om=0} =\fs_0(x^0)= 1/(1+2\pi\la e^{x^0})$. Then the integrand of the first term has a simple form and oscillates for $x^0 \to -\infty$ while that of the second one is complicated but vanishes exponentially in both directions $x^0 \to \pm\infty$. Hence the $\delta$ contribution comes solely from the first term which can be integrated exactly,
while for the second integral is well defined even for imaginary $\xi_1+\xi_2$ and the analytic continuation of \eq{Aguess} from the region of $\mathrm{Re} (\xi_1+\xi_2)>0$ can be trusted. The $\delta$-term that adds to \eq{Aguess} is seen to be\footnote{Naturally, a similar term also appears in the exact amplitude, and is given by eliminating the superscript ``$\appro$'' of $I$ in \eq{deltat}.}
\be \label{deltat}
 \As_{2,\delta} = \pi \delta_{0,\vek_1+\vek_2} I^\appro_{i\om_1,i\om_2}(N=0) \delta(\om_1+\om_2) \ .
\ee

Naively, since we effectively replace $N \to -i(\om_1+\om_2)$ in the asymptotic expansion \eq{gasguess}, one would expect the improved asymptotic formula \eq{Aguess} to be a good estimate for large energies for which $1/N$ is small. Unfortunately, the correction to \eq{gasguess} is likely to include terms which are $\sim \om_a/N$ (or worse) and become (at least) $\CO(1)$ at $N \sim \om_a$. Thus the result \eq{Aguess} only serves as a model of the exact result for any values of the energies. However, note that the calculations of Appendix~\ref{exampapp} suggest that the correction terms are small:
the improved approximation is seen to work well also for small values of $N$ and also when continued to negative (but small) values of $N$.

It is straightforward to check that the analysis of Subsection~\ref{sec:contour} can be similarly applied to the approximate asymptotic formula \eq{gasguess} for $n>2$.
The consequent generalization of \eq{Aguess} can be simplified to read
\be \label{Anfres}
 \As_n \approx \delta_{0,\sum_a \vek_a} {\cal N}\left[(\vek_a)\right] (2\pi\la)^{-i\sum_a\om_a} \Gamma\left(i\sum_a\om_a\right)\prod_{a=1}^n\frac{G(i\om_a+1)^3G(2-i\om_a)}{G(2i\om_a+1)}\ .
\ee
As discussed in Subsection~\ref{sec:asympt} the $\vek_a$ dependence factorizes into the factor ${\cal N}\left[({{\vek}_a})\right]$ at least for small $\vek_a$.
We give the singularities and the asymptotic behavior of the model amplitude in Appendix~\ref{sec:zeros}.

Notice that our approximation is not restricted to only positive integer valued $\xi_a$. The original asymptotic formula \eq{asres} is valid (see \cite{Basor}) for $\mathrm{Re}\,{\xi_a}>-1/2$. While our improved formula \eq{gasguess} was motivated using integer $\xi_a$, it approaches the original one \eq{asres} at $N \to \infty$ for any set of (complex) $\xi_a$'s. In Appendix~\ref{exampapp} it was demonstrated that the improved asymptotics works much better than \eq{asres} for sets of integer $\xi_a$.  There is no reason to believe why it should fail to be an improvement also for imaginary $\xi_a = i \omega_a$.

{\em Final comments}. We conclude with some final thoughts. We have presented a method to calculate $n$-point boundary functions. It would be important to
develop similar methods for $n$-point bulk correlators. The main obstacle for a straightforward generalization of our calculations is the following. The boundary
operators correspond to test charges that we constructed from unit charges of the Dyson gas. However, the bulk operators cannot similarly be made of the unit charges
on the boundary -- a different trick must be found for the bulk correlation function calculations.  Another important issue is to develop a clear estimate how
good an estimate (\ref{nptfinal}) is for the amplitude. A promising way to test our method would be to work directly in spacelike boundary Liouville theory,
use our method to compute the boundary two-point function, and then compare with the exact known result of \cite{Fateev:2000ik,Teschner:2000md,Ponsot:2001ng}.


\bigskip
\bigskip

\noindent

{\bf \large Acknowledgments}

\bigskip

We thank A. Dabholkar, D. D. Dietrich, M. Gaberdiel, P. Kraus, F. Larsen, J. Majumder, S. Murthy, and A. Sen for useful discussions and the anonymous JHEP referee for useful comments. N.J. thanks the hospitality of the University of Southern Denmark during the preparation of this article. N.J. has been supported in part by the Magnus Ehrnrooth Foundation. M.J. has been supported in part by the Magnus Ehrnrooth Foundation and by the Marie Curie Excellence Grant under Contract No. MEXT-CT-2004-013510.  This work was also partially supported by the EU 6th Framework Marie Curie Research and Training Network ``UniverseNet'' (MRTN-CT-2006-035863).

\bigskip
\bigskip

\appendix


\section{On KRS relation}\label{sec:appKRS}

In this Appendix we check \eq{roltachfor}.
Following \cite{Kraus:2002cb}, equation (4.7), we need
to calculate
\beqa \label{Kraus}
\mathcal{B} & = & \int_\mathrm{disk} d^2w \left[\frac{w^2-e^{2i\tau}}{2e^{i\tau}}\langle \partial X^0(w,\bar w) e^{i\omega X^0(\tau)}e^{-S_\mathrm{bdry}}\rangle'\right.\nn\\
            &   & +\left.\frac{\bar w^2-e^{-2i\tau}}{2e^{-i\tau}}\langle \bar \partial X^0(w,\bar w) e^{i\omega X^0(\tau)}e^{-S_\mathrm{bdry}}\rangle'\right] \ ,
\eeqa
where the primes of the expectation values indicate that the zero mode $x^0$ is left unintegrated. We start from (straightforward use of Wick theorem)
\beqa
 \mathcal{C}
  & = & \langle e^{i\omega_cX^0(w,\bar w)} e^{i\omega X^0(\tau)}\prod_{i=1}^N e^{X^0(t_i)} \rangle' \\\nn
  & = & \left|1-w\bar w\right|^{-\omega_c^2/2}\left|w-e^{i\tau}\right|^{-2\omega\omega_c}\prod_{i<j}\left|e^{it_i}-e^{it_j}\right|^2 \prod_i\left|e^{i\tau}-e^{it_i}\right|^{2i\omega}\left|w-e^{it_i}\right|^{2i\omega_c}
\eeqa
which is to be integrated over $t_i=0 \ldots 2\pi$ and summed over $N$. Notice that
\beqa
 &   & \langle \partial X^0(w,\bar w) e^{\xi X^0(\tau)}\prod_i  e^{X^0(t_i)} \rangle' = -i\left.\frac{\partial^2}{\partial w \partial \omega_c} \mathcal{C}\right|_{\omega_c=0} \nn\\
 & = & \left[\sum_i\inv{w-e^{it_i}}+\frac{\xi}{w-e^{i\tau}}\right] \prod_{i<j}\left|e^{it_i}-e^{it_j}\right|^2 \prod_i\left|e^{i\tau}-e^{it_i}\right|^{2\xi}
\eeqa
and similarly for the term containing $\bar\partial$ in \eq{Kraus}. Recall that $\xi=i\omega$. Let us do the $w$ integration first. The $w$-dependent part reads
\beq
 I_w = \int d^2 w\left[ \frac{w^2-e^{2i\tau}}{2e^{i\tau}}\left(\sum_i\inv{w-e^{it_i}}+\frac{\xi}{w-e^{i\tau}}\right) + \mathrm{h.c.}\right] \ ,
\eeq
where the $\mathrm{h.c.}$ assumes real $\xi$. Developing the integrand at $w, \bar w=0$ we see that only the constant term survives,
\beq
 I_w = \pi \xi + \frac{\pi}{2}\sum_i\left(e^{i\tau-it_i}+e^{-i\tau+it_i}\right) = \pi\left[\xi + \sum_i\cos(\tau-t_i)\right] \ .
\eeq
Let us then do the $t_i$ integrations. The integral $\propto \xi$ is the $I_\xi(N)$ discussed above, and for the second term we use \eq{gasintegral}.
Putting the results together,
\beqa
 \mathcal{B}
  & = & \pi e^{\xi x^0} \sum_N (-z)^N \left[\xi-\frac{N\xi}{N+\xi}\right] I_\xi(N) = \pi \xi^2 e^{\xi x^0} \sum_N \frac{(-z)^N}{N+\xi} I_\xi(N) \nn\\
  & = &
  \frac{\pi \xi^2 }{z^\xi}\int_0^z dz' (z')^{\xi-1} \fs_1(z') \ .
\eeqa
Moreover, the $x^0$ dependencies of the $1/z^\xi$ and the $e^{\xi x^0}$ exactly cancel, whence after derivating with respect to $x^0$ \cite{Kraus:2002cb}
\beq
 \frac{\partial \mathcal{B}}{\partial x^0} = \pi \xi^2  \fs_1 \ .
\eeq
We have thus checked the formula (3.14) of \cite{Kraus:2002cb} in this special case. The $\xi^2$ in the proportionality constant arises from the conformal dimension of the operator $e^{\xi X^0}$, which is included in the normalization factor ${\cal \widetilde N}$ of Subsection~\ref{sec:onepoint}.


\section{A special case of the $n$-point amplitude}\label{exampapp}

To clarify the involved derivation of the model for the $n$-point amplitude in Section~\ref{npointsec}, we consider here the simplest nontrivial example, $(\xi_1,\xi_2)=(2,2)$ and ${\vec k}_1\cdot {\vec k}_2=0$,\footnote{Notice that the condition ${\vec k}_1\cdot {\vec k}_2=0$ eliminates all dependence on spatial momentum. It actually conflicts with momentum conservation.} that can be calculated also exactly. We have also checked other cases of sets of small integers, and found similar results.

Let us start with the results for the integral $I_{2,2}(N)$ of \eq{Indef} which equals $I_2(N,2)/N!$ of \cite{Jokela:2007dq}. Hence we have
 \bea \label{I22form}
  I_{2,2}(N) &=& \frac{2}{8!}\left[35\frac{(N+8)!}{N!}+77\frac{(N+7)!}{(N-1)!}+27\frac{(N+6)!}{(N-2)!}+\frac{(N+5)!}{(N-3)!}\right] \nn\\
             &=& \frac{35 N^3+ 467 N^2+ 2046 N+2940  }{5040}\prod_{k=1}^5(N+k) \nn\\
   I^\asymp_{2,2}(N) &=& \left[\frac{(N+3)(N+2)^2(N+1)}{12}\right]^2 = \frac{(N+3)^2(N+2)^4(N+1)^2}{144}\nn\\
  I^\appro_{2,2}(N) &=& \frac{(N+2)(N+1)}{(N+4)(N+3)}\left[\frac{(N+5)(N+4)^2(N+3)}{12}\right]^2 \nn\\
  &=& \frac{(N+5)(N+4)^2}{144}\prod_{k=1}^5(N+k)
\eea
where the first, the second, and third formula are the exact result, the one obtained from the asymptotic formula \eq{asres}, and the improved asymptotic formula \eq{gasguess}, respectively. For $N \to \infty$ we find
\bea
 \frac{I^\asymp_{2,2}(N)}{I_{2,2}(N)} &=& 1 -\frac{432}{35 N} + \frac{142384}{1225 N^2} + \morder{\inv{N^3}} \nn\\
  \frac{I^\appro_{2,2}(N)}{I_{2,2}(N)} &=& 1 -\frac{12}{35 N} + \frac{2594}{1225 N^2} + \morder{\inv{N^3}} \ .
\eea
Some of the values of the integrals are tabulated in Table~\ref{inttab}. The improved formula works much better, in particular for low values of $N$. For higher values of $\xi_a$ the improvement is even more drastic, basically since
the difference between the  effective number of unit  charges $\tilde N = N + \sum_{a=1}^n\xi_a$, which is used in the improved asymptotics, and the actual number of unit charges $N$ increases.

\begin{table}\caption{Exact and approximated values of the integral $I_{2,2}(N)$. The first tabulated row is the exact result of the integral, while the two others are given by the asymptotic formulae \eq{asres}, \eq{gasguess}, written explicitly in \eq{I22form}.}\label{inttab}
\begin{tabular}{|c||c|c|c|c|c|c|c|c|}
 \hline
  $N$ & 0 & 1 & 2 & 3 & 5 & 10 & 100 \\
\hline\hline
 $I_{2,2}(N)$ & 70& 784& 4590& 18968& 175320& 7514650&  91680976745020 \\
 $I^\asymp_{2,2}(N)$ & 1& 36& 400& 2500& 38416& 2944656&  81349594398801 \\
 $I^\appro_{2,2}(N)$ & 200/3& 750& 4410& 54880/3& 170100& 7357350&  91384995374400 \\
\hline
\end{tabular}
\end{table}

The two-point function
\be
 \fs_2(x^0)\big|_{\xi_1=\xi_2=2} =\sum_{N=0}^\infty (-z)^N I_{2,2}(N)\ ,
\ee
where $z=2\pi \la e^{x^0}$, can be calculated explicitly for all the results \eq{I22form}. In particular, for large $x^0$ we have
\bea  \label{gasymp}
 \fs_2(x^0) &=& - \frac{2}{(2\pi \la)^6}e^{-6x^0} + \frac{72}{(2\pi \la)^7}e^{-7x^0} +\morder{e^{-8x^0}} \nn\\
 \fs_2^\asymp(x^0) &=& - \frac{1}{(2\pi \la)^4}e^{-4x^0} + \frac{36}{(2\pi \la)^5}e^{-5x^0} +\morder{e^{-6x^0}} \nn\\
\fs_2^\appro(x^0) &=& - \frac{10}{3(2\pi \la)^6}e^{-6x^0} + \frac{90}{(2\pi \la)^7}e^{-7x^0} +\morder{e^{-8x^0}} \ .
\eea
The improved asymptotic formula produces also the large $x^0$ asymptotics nicely: $\fs_2^\appro$ is correct up to the proportionality constant for $x^0 \to \infty$.

The analytic continuation of $I_{2,2}(N)$, $J_{2,2}(s)$, is found by letting $N \to -s$ in \eq{I22form}. The result for the integrated amplitude is then obtained by applying \eq{Aguess} to the three different cases of \eq{I22form}, which  gives
\be
 \As_2\big|_{\xi_1=\xi_2=2}= \lim_{s\to 4}\frac{\pi(2\pi\la)^{-4}}{\sin\pi s}   J_{2,2}(s)=-\frac{1}{70(2\pi\la)^4}\, ,\ \infty\, , \ 0
\ee
where the first, the second, and the third numbers are the exact result, the result for the naive asymptotic formula \eq{asres}, and the result for the improved formula \eq{gasguess}, respectively. Both the asymptotic approximations give an incorrect result by an infinite factor. However, the numerical factor $-1/70$ of the exact result is extremely small when compared, \eg, to the series coefficients of Table~\ref{inttab}, whence the zero result obtained by the improved asymptotic formula should be, in fact, considered as a good approximation. It would be interesting to be able to compare our model amplitude to the exact result for more physical, noninteger values of $\xi_a$ where no accidental zeroes or infinities are expected to occur.

We end this Appendix by an encouraging observation in a more general setup. It is straightforward to check that, in fact, for any sets of integer $(\xi_1,\ldots \xi_n)$ the expected exact asymptotic behavior \cite{Jokela:2007dq}
\be
 \fs_n(x^0) \sim \exp\left[-x^0 \sum_a\xi_a-x^0 \max\{\xi_a\} \right] \ ,
\ee
is reproduced by $\fs_n^\appro$ similarly as for $(\xi_1,\xi_2)=(2,2)$ in \eq{gasymp}. We denote the analytic continuation of $I^\appro_{\xi_1,\ldots \xi_n}(N)$ of \eq{gasguess} to complex $s=-N$ by $J^\appro_{\xi_1,\ldots \xi_n}(s)$ in analogue to \eq{Jdef}, \eq{Jappdef} above.
Using the behavior of Barnes $G$ near its zeroes from Appendix~\ref{sec:zeros}, a lengthy calculation shows that the first pole of  $J^\appro_{\xi_1,\ldots \xi_n}(s)/\sin \pi s$ on the positive real axis occurs at $s=\sum_a\xi_a+\max_a\{\xi_a\}$ in the special case of integer $\xi_a$. Hence, for the $n$-point function and integer $\xi_a$, \eq{lzas} indeed becomes
\be
 \fs^\appro_n(x^0) \sim \exp\left[-x^0 \sum_a\xi_a-x^0 \max\{\xi_a\} \right] \ .
\ee
In other words, the corresponding poles of $J^\appro_{\xi_1,\ldots \xi_n}$ and the exact analytic continuation $J_{\xi_1,\ldots \xi_n}$ coincide. Note that these poles lie at positive $s$, \ie, negative $N$, while $J^\appro_{\xi_1,\ldots \xi_n}$ results from an asymptotic formula \eq{gasguess} for large positive $N$. This observation gives more confidence to the model amplitude of \eq{Aguess}, \eq{Anfres} which was derived using \eq{gasguess}.


\section{Singularities and asymptotics of the model amplitude}\label{sec:zeros}

Barnes $G(z)$ is an entire function and has its zeroes on the
negative real axis,
\beq \label{Gzeros}
 G(z+1) = (-1)^{n(n-1)/2} G(n+1)(z+n)^n \left[1+\CO\left(z+n\right)\right] \ ,
\eeq for $n=1,2,\ldots$. Hence all the special points (zeroes or
singularities) of the one-point amplitude $\hat \As_1(\om)$ of \eq{1ptres} lie on the imaginary axis, \beqa
 \hat \As_1 &=& \frac{(-1)^{n(n-1)/2}(2\pi\la)^{-n} G(n+1)^4}{G(2n+1)}\, (i\om-n)^{n-1}\left[1+\CO\left(i\om-n\right)\right]\ ;\nn\\
 \hat \As_1 &=& \frac{(-1)^{n(n-1)/2}(2\pi\la)^{n} G(n+1)^4}{2^{2n}G(2n+1)}\, (i\om+n)^{n-1}\left[1+\CO\left(i\om+n\right)\right]\ ;\nn \\
 \hat \As_1 &=& -\frac{\pi(2\pi\la)^{n+1/2} G(1/2-n)^3 G(3/2+n)}{2^{2n+1}G(2n+2)}\, \nn\\
     &&\times (i\om+n+1/2)^{-2n-1}\left[1+\CO\left(i\om+n+1/2\right)\right]
\eeqa
for any $n=0,1,2,\ldots$. In particular, poles are found at
$\om=0$ (where $ \hat \As_1 \sim 1/i\om$) and at $\om = i/2,
3i/2,5i/2,\ldots$.

The singularities of the model $n$-point amplitude
\be  \label{nmod}
 \hat \As_n \approx (2\pi\la)^{-i\sum_a\om_a} \Gamma\left(i\sum_a\om_a\right)\prod_{a=1}^n\frac{G(i\om_a+1)^3G(2-i\om_a)}{G(2i\om_a+1)}
\ee
arise similarly from the poles of $\Gamma\left(i\sum_a\om_a\right)$ and from the zeroes of each $G(2i\om_a+1)$. As above, the hat denotes that we dropped the $\vek_a$ dependent terms. At the possible singularities $\sum_a\om_a \simeq -m$, $i\om_b \simeq - m$, $i\om_b \simeq  m+1$, and $i\om_b\simeq -m-1/2$, we find
\bea
 \hat \As_n &=& \frac{(-2\pi \la)^m}{m!}\prod_{a=1}^n\frac{G(i\om_a+1)^3G(2-i\om_a)}{G(2i\om_a+1)}\bigg|_{i\sum_a\om_a=-m} \\
 &&\times\frac{1}{i\sum_a\om_a+m}\left[1+\morder{i\sum_a\om_a+m}\right]\ ;\nn\\
 \hat \As_n &=& \frac{(-1)^{m(m+1)/2}(2 \pi\la)^{m}(2\pi\la)^{-i\sum_{a \ne b}\om_a} \Gamma(i\sum_{a \ne b} \om_a - m)G(m+1)^3G(2+m)}{2^{2m}G(2m+1)} \nn \\
     &&\times\prod_{a\ne b} \frac{G(i\om_a+1)^3G(2-i\om_a)}{G(2i\om_a+1)}\times ( i\om_b + m)^{m}\times\left[1+\CO\left( i\om_b+ m\right)\right]\ ; \nn \\
 \hat \As_n &=& \frac{(-1)^{m(m+1)/2}(2\pi\la)^{-i\sum_{a \ne b}\om_a} \Gamma(i\sum_{a \ne b} \om_a + m+1)G(m+2)^3G(m+1)}{(2\pi\la)^{m+1}G(2m+3)} \nn \\
     &&\times\prod_{a\ne b} \frac{G(i\om_a+1)^3G(2-i\om_a)}{G(2i\om_a+1)}\times ( i\om_b - m-1)^{m}\times\left[1+\CO\left( i\om_b- m-1\right)\right]\ ; \nn \\
 \hat \As_n  &=& \frac{(-1)^m(2\pi\la)^{m+1/2-i\sum_{a \ne b}\om_a} \Gamma(i\sum_{a \ne b} \om_a-m-1/2) G(1/2-m)^3 G(5/2+m)}{2^{2m+1}G(2m+2)} \nn\\ \nn
     &&\times\prod_{a\ne b} \frac{G(i\om_a+1)^3G(2-i\om_a)}{G(2i\om_a+1)}\times (i\om_b\!+\!m\!+\!1/2)^{-2m-1}\times\left[1\!+\!\CO\left(i\om_b\!+\!m\!+\!1/2\right)\right]
\eea
respectively,
for each $m=0,1,2,\ldots$ and $b=1,2,\ldots,n$, and assuming that the singularities are distinct (which requires $n>1$). We thus find poles only at negative integer values of $i\sum_a\om_a$ and at negative half-integer values of each $i\om_b$ whereas the amplitude vanishes for almost all integer values of each $i\om_b$.

Since the model amplitude has a factorized form, its asymptotics can be immediately written down based on the one-point 
result  \eq{1ptasympt}. It reads
\bea
\hat \As_n &=& (2\pi\la)^{-i\sum_a\om_a} \Gamma\left(i\sum_{a=1}^n\om_a\right) \prod_{a=1}^n\Gamma(1-i\om_a) \nn\\
         &   &\times
\exp\Bigg\{\sum_{a=1}^n\left[\om_a^2\left(\frac{i\pi}{2}\mathrm{sgn}\left(\mathrm{Re}\,\omega_a\right) +2\log 2\right)
 -\frac{1}{4}\log\left(i\om_a\right) \right. \nn\\
& &\left. -\frac{i\pi}{12}\mathrm{sgn}\left(\mathrm{Re}\,\omega_a\right)+\frac{1}{12}\log 2+3\zeta'(-1) \right]\Bigg\} \prod_{a=1}^n\left[1+\CO\left(\om_a^{-2}\right)\right]
\ . \label{nptfinal}
\eea
Note that the prefactor vanishes exponentially for any $\omega_a \to \infty$ and thus does not affect the leading asymptotic behavior $\sim \exp[\om_a^2]$, which is similar to that of the one-point amplitude in \eq{1ptasympt}.


\end{document}